# Strain induced band gap deformation of H/F passivated graphene and h-BN sheet

Authors: A. Bhattacharya[1], S. Bhattacharya[1], G. P. Das[1*]

Department of Materials science, Indian Association for the Cultivation of Science, Jadavpur, Kolkata 700032, INDIA.

*Corresponding Author's email: msgpd@iacs.res.in
Authors' email address: msab2@iacs.res.in, mssb3@iacs.res.in

**Abstract:**

Strain induced band gap deformations of hydrogenated/fluorinated graphene and hexagonal BN sheet have been investigated using first principles density functional calculations. Within harmonic approximation, the deformation is found to be higher for hydrogenated systems than for the fluorinated systems. Interestingly, our calculated band gap deformation for hydrogenated/fluorinated graphene and BN sheets are positive, while those for pristine graphene and BN sheet are found to be negative. This is due to the strong overlap between nearest neighbor $\pi$ orbitals in the pristine sheets, that is absent in the passivated systems. We also estimate the intrinsic strength of these materials under harmonic uniaxial strain, and find that the in-plane stiffness of fluorinated and hydrogenated graphene are close, but larger in magnitude as compared to those of fluorinated and hydrogenated BN sheet.

**Manuscript Text:**

Graphene[1], the two dimensional sp$^2$ bonded single layer of graphite, is a hot pursuit today because of its unusual electrical[2] and mechanical properties[3,4] and their implementation in future applications. Recently, hydrogenated derivative of graphene called graphane has been predicted theoretically[5,6], followed by it's synthesis in laboratory by Geim's group[7]. Subsequently, various interesting properties of this material such as reversible hydrogenation-dehydrogenation[8], insulating wide band gap[5-7], vacancy generated quantum dots[8] etc were revealed. The conductivity of graphene being very high (its electrons acting as massless fermions), its usage in electronic devices can be made possible by introducing a band gap[9]. Hydrogenation of graphene is one such option for opening of a band gap and hence might be useful for band gap engineering[10] in graphene-based devices. Another analogous two dimensional nanostructure viz. hexagonal Boron Nitride sheet[11] (h-BN sheeet) has emerged as a strong candidate for possible modification of the electronic structure of graphene. BN sheet has recently been experimentally synthesized in single and multiple layers[12,13]. While graphene is a semi-metal with unique 2D conducting properties, h-BN is an insulator with wide band gap of ~7 eV. Hydrogenation of graphene leads to widening of band gap, while hydrogenation of BN-sheet leads to reduction in the band gap of the sheet[14-16]. Similar results can be realized in case of fluorination in both graphene and BN-sheet[14]. Hydrogenation/fluorination of graphene/BN sheet takes place in such a way that one H/F atom gets bonded to each of the C/B/N atoms of the sheet in specific periodic fashion, giving rise to various possible conformers of the sheet viz. chair, boat and stirrup. In the chair and boat conformers, H/F atom alternates singly and in pair on both sides of the sheet[5,14,15], while in the stirrup conformer, three consecutive H/F atoms alternate on either side of the sheet[16,17]. In CH, CF and BFNF sheets, the chair conformer has been found to have the highest stability. However, only in case of BHNH sheet, the stirrup conformer is found to have the highest stability with binding energy of ~4.84 eV/atom followed by boat and chair conformer[16]. The tuning of band gap in graphene/h-BN by chemical functionalization of various dopant is a natural means to modify the band gap in wider range and therefore, it has recently been explored by various groups[18,19]. The local deformation in the crystal lattice due to increase in mechanical strain can also result in change in the effective electric potential experienced by free electrons in a semiconductor which in turn leads to increase/decrease in the band gap of the crystal[20-22]. Recently, the elastic properties and intrinsic strength of graphene monolayer has been measured by Lee et.al.[3] using nanoindentation experiments on graphene membranes. Similar measurement can be carried out on stable 2D analoges of graphene such as graphane[5], BN sheet[11], BHNH sheet[16] etc. Thus, a systematic study of band gap deformation and elastic properties of these planar nanostructures with increase in strain on various chemically modified graphene and graphene like systems is very relevant in this context.

In this work, we use first principles approach to study the band gap deformation of hydrogenated/fluorinated graphene and BN sheet under homogeneous (biaxial) strain applied within the harmonic limits. We find that the strain induced band gap deformation is higher for hydrogenation as compared to fluorination for both graphene and BN sheet. Interestingly, our calculated band gap deformation for hydrogenated/fluorinated graphene and BN sheets are positive, while those for pristine graphene and BN sheet are found to be negative. We also estimate the intrinsic strength of these materials under harmonic uniaxial strain. We find that the in-plane stiffness of fluorinated and hydrogenated graphene are close, but larger in magnitude as compared to those of fluorinated and hydrogenated BN sheet. On comparing the in-plane stiffness of these H/F passivated planar sheets with those of pure graphene and BN sheet, it is found that the parent sheets have higher intrinsic strength as compared to their hydrogenated and fluorinated counterparts.

We have performed first principles density functional (DFT)[23,24] based all electron calculations using DMol3 codes[25] with a double numerical quality basis set under a polarization function (DNP). The calculations have been carried out under generalized gradient approximation (GGA) using the PW91 exchange-correlation functional of Perdew et al[26,27]. An energy cut off of 400eV has been used. The K-mesh [12X12X4] is generated by Monkhorst–Pack[28] method, and all results are tested for convergence with respect to mesh size. In all the calculations, the 2D sheet is placed in a vacuum of 12 Å, in order to hinder the interaction between two adjacent layers. All our calculations have been checked for self-consistency with convergence to $1 \times 10^{-6}$ eV in total energy.

A chair graphane layer has hexagonal unit cell ($a_{xo} = a_{yo}$, $\alpha = 90^\circ$, $\beta = 90^\circ$, $\gamma = 120^\circ$) with an equilibrium lattice parameter of 2.54 Å and the equilibrium C-H and C-C bond lengths are found to be 1.11 Å and 1.54 Å respectively (Table-1). We first study the behavior of graphane under homogeneous-biaxial strain applied in the harmonic region with $\Delta a_x/a_{xo} = \Delta a_y/a_{yo} = \pm 0.02$. The variation of binding energy (BE) per atom with the normalized volume expansion of lattice ($V/V_o$, where $V_0$ is the equilibrium volume of the unit cell) is shown in the Fig-1(a) of SI-1. The BE versus normalized volume curve is a parabola in the range $\Delta V/V_0 = \pm 0.03$, with minimum at the equilibrium cell volume (ie. $V=V_0$), increasing on either sides corresponding to the compression ($-V/V_0$) or expansion ($+V/V_0$). The unstrained equilibrium point has the highest binding energy of ~5.227 eV/atom and a band gap of 4.62 eV. Similar calculations have also been performed for the fluorinated counterpart of graphane having a formula unit of CF. The equilibrium lattice parameter of chair CF sheet is found to be 2.6Å within the symmetric constraint of hexagonal lattice (ie. $a_{xo}= a_{yo}$). The equilibrium C-F and C-C bond lengths in a hexagonal CF lattice are estimated as 1.37Å and 1.58Å respectively (Table-1). The CF sheet has an equilibrium BE of 5.467 eV/atom [See Fig-1(b) of SI-1] which is slightly higher (by 0.24 eV/atom) than that of graphane. However, the band gap of fluoro-graphene sheet (3.45 eV) is found to be lower than that of graphane by ~1.17 eV.

We now discuss the effect of homogeneous biaxial compression and expansion (within harmonic limits) on hydrogenated BN-sheet. We study the effect of strain on the chair conformer of hydrogenated BN sheet (BHNH sheet) which also has a hexagonal unit cell with lattice parameters close to those of graphane. The equilibrium lattice parameters are found to be 2.59 Å, which corresponds to the minimum of the BE Vs $V/V_0$ parabola [See Fig-1(c) of SI-1]. The equilibrium B-H, N-H and B-N bond lengths are 1.2 Å, 1.03 Å and 1.58 Å respectively. The stability of the hydrogenated BN sheet has been found to be ~ 4.59 eV/atom, while the equilibrium band gap is estimated as 4.46 eV. Extending similar sets of calculation for the fluorinated BN sheet, we find that the chair conformer of fluorinated BN sheet (BFNF sheet) has an equilibrium lattice parameter of 2.67 Å. As observed in case of both graphene and BN sheet, lattice parameter increases upon fluorination as compared to hydrogenation (Table-1). The equilibrium B-F, N-F and B-N bond lengths are 1.35 Å, 1.44 Å and 1.64 Å respectively. Unlike chair BHNH sheet, the N-F bond length is higher than the B-F bond length in the BFNF sheet. The stability of fluorinated BN sheet (4.89 eV/atom) is found to be higher than its hydrogenated counter part while the band gap (3.59 eV) is lower for former. Thus, fluorination reduces the band gap as compared to hydrogenation in both graphene and BN sheet which is due to the increase in BN/C-C bond on fluorination as compared to that on hydrogenation (Table-1). It is imperative to compare the effect of harmonic biaxial strain/stress on the band gap deformation in these hydrogenated and fluorinated sheets. The variation of band gap deformation with the increase of biaxial strain in CH, CF, BN, BHNH and BFNF systems are given in Fig-1. The band gaps of these hydrogenated and fluorinated graphene/BN sheet increase linearly with expansion of lattice which is similar in case of convential bulk semiconductors such as Si, GaAs etc[21,22]. The slope of $E_g$ versus $\ln(V/V_0)$ plot gives us the estimation of band gap deformation potential ($D_P$) with the change in volume of the unit cell [$dE_g/d(\ln V)$]. $D_P$ is found to be highest for graphane (9.07 eV/Å) while it is the lowest for fluoro-graphene (0.88 eV/Å). In hydrogenated BN sheet $D_P$ (6.61 eV/Å) is also estimated to be higher than the fluorinated BN sheet (1.82 eV/Å). The trend can be summarized as follows (see Table-1):

$$[dE_g/d(\ln V)]_{CH} > [dE_g/d(\ln V)]_{BHNH} > [dE_g/d(\ln V)]_{BFNF} > [dE_g/d(\ln V)]_{CF}$$

Therefore, in both graphene and BN sheet, the $D_P$ on hydrogenation is higher than that on fluorination. We have also compared these results with the band gap deformation in graphene and BN sheet, in both cases lattice stretching leads to decrease in band gap. The $D_P$ in BN sheet is estimated to be -3.88 eV/Å, which is lower in magnitude than that of its hydrogenated counter part. Graphene also has negative band gap deformation but its deformation is much lower in magnitude (-0.42 eV/Å) than both of its hydrogenated and fluorinated counter parts (Table-1). The negative slope of band gap deformation in graphene and BN sheet is due to the strong overlapping between nearest neighbor π orbitals in the sheets (the π-π interaction with change in inter-atomic distance differs from the σ-σ interaction). However, in case of their hydrogenated/fluorinated counter parts, functionalization leads to passivation of these dangling π bonds due to $sp^3$ hybridization. Therefore, in these hydrogenated/fluorinated sheets the conduction of electron depends solely on the C-C/B-N bonding. Upon expansion of the cell parameters the C-C/B-N bonding becomes weak and hence leads to decrease in conduction and increase in band gap in these passivated systems.

Now we discuss Young's modulus which dictates the intrinsic strength of any homogeneous isotropic three dimensional crystals. In case of two dimensional planar nanostructures under consideration here, the thickness of the layer is ambiguous because of their reduced dimensionality and therefore it makes more sense in calculating the in-plane stiffness (I) of the materials. The in-plane stiffness of the materials can be expressed as $I = 1/A_0 \, (\partial^2 E_S / \partial S^2)$[19], where $A_0$, $E_s$ and S are the equilibrium area, the strain energy (difference between the energy of the strained system and equilibrium energy) and the uniaxial strain ($\Delta a_y/a_{yo}$) respectively. In order to determine 'I' of the systems, we keep $a_x$ fixed (as the equilibrium lattice parameter, $a_{xo}$) while $a_y$ is varied in the harmonic region (i.e. $\Delta a_y/a_{yo} = \pm\, 0.02$). The response of strain energy ($E_S$) for hydrogenated and fluorinated graphene and BN systems under the uniaxial strain along Y-axis is shown in Fig-2. The $E_S$ versus S curves are again parabolic with their vertex at the equilibrium point ($a_{xo}$, $a_{yo}$). The strain energy increases with increase in strain from the equilibrium position and the slope of first derivative of the $E_S$ vs S curve (Table-2) is the measure of the 'in plane stiffness' of the material (Fig-3). The 'I' of graphane (246.55 J/m$^2$) and fluoro-graphane (255.55 J/m$^2$) are found to be close but larger in magnitude to those of hydrogenated BN sheet (181.56 J/m$^2$) and fluorinated BN sheet (170.82 J/m$^2$). Comparing our calculated value of 'I' with the estimated value of 'I' in Ref. 19 for graphene, graphane and BN sheet, we find reasonably good agreement. Thus, the in-plane stiffness of nano-sheets which are chemically derived from graphene are found to be higher compared to those derived from BN-sheet. This is also evident from the higher stabilities of graphane and fluoro-graphane than hydrogenated and fluorinated BN-sheet. We have also compared the in plane stiffness of these hydrogenated and fluorinated sheets with those of native graphene and BN sheet. We find that graphene has the highest value of 'I' of 358.5 J/m$^2$ which is followed by that of BN sheet having 'I' of 281.6 J/m$^2$. Therefore, upon hydrogenation/fluorination, the inherent strength of these sheets decreases as compared to their parent sheets. Our estimated result supports the experimental value of I, as measured by nanoindentation of freestanding graphene membrane by Lee et.al[3]. Thus, the order of 'in plane stiffness' can be summarized as below;

$$I_{Graphene} > I_{BN} > I_{CF} > I_{CH} > I_{BHNH} > I_{BFNF}$$

In summary, we have compared the band gap deformation and mechanical properties of hydrogenated/fluorinated graphene and BN sheet under homogeneous (biaxial) and uniaxial strain, using first principles based density functional theory. We find that the homogeneous biaxial strain induced band gap deformation is higher for hydrogenation as compared to that for fluorination for both graphene and BN sheet. We have also estimated the in plane stiffness of these H/F passivated planar sheets and compared the values with those of graphene and BN sheet. We have found that graphene has the highest in plane stiffness which is followed by BN sheet. Upon hydrogenation/fluorination the inherent strength of these passivated sheets decreases as compared to their parent counterparts. The in-plane stiffness of fluorinated and hydrogenated graphene are found to be close, but higher than that of fluorinated and hydrogenated BN sheet.

Table-1: Calculation of band gap deformation potential ($D_P$) of Graphene, CH, CF, BN, BHNH and BFNF sheets under homogeneous biaxial strain applied within the harmonic limits (negative sign implies band gap decrease with lattice expansion)

| System | Eq. Bond Length (Å) | | Eq. lattice parameter (Å) $a_o$ | Eq. Lattice Volume (Å$^3$) $a_o^2 \times 12$ $V_0$ | BE (eV/atom) | Band Gap (eV) | $dE_g/d(\ln V)$ (eV/Å) '$D_P$' |
|---|---|---|---|---|---|---|---|
| Graphene | C-C | 1.42 | 2.46 | 72.62 | -7.952 | 0 | -0.42 |
| CH | C-C | 1.54 | 2.54 | 77.42 | -5.227 | 4.62 | 9.07 |
|  | C-H | 1.11 | | | | | |
| CF | C-C | 1.58 | 2.6 | 81.12 | -5.467 | 3.45 | 0.88 |
|  | C-F | 1.37 | | | | | |
| BN | B-N | 1.45 | 2.52 | 76.20 | -7.071 | 4.68 | -3.88 |
| BHNH | B-N | 1.58 | 2.59 | 80.50 | -4.591 | 4.46 | 6.61 |
|  | B-H | 1.20 | | | | | |
|  | N-H | 1.03 | | | | | |
| BFNF | B-N | 1.62 | 2.67 | 85.55 | -4.894 | 3.59 | 1.82 |
|  | B-F | 1.35 | | | | | |
|  | N-F | 1.44 | | | | | |

Table-2: Estimation of 'in plane stiffness (I)' of Graphene, CH, CF, BN, BHNH and BFNF sheets under uniaxial strain along Y axis;

| System | $d^2E_s/dS^2$ in (eV) | $I = 1/A_0(d^2E_s/dS^2)$ in (J/m$^2$) |
|---|---|---|
| Graphene | 135.413 | 358.47 |
| CH | 99.258 | 246.55 |
| CF | 107.825 | 255.55 |
| BN | 111.622 | 281.58 |
| BHNH | 76.016 | 181.56 |
| BFNF | 76.994 | 170.82 |

Figure caption (color online);

Fig-1: Comparison of band gap variation with the normalized area under homogeneous biaxial strain in hydrogenated and fluorinated graphene and BN sheet.

Fig-2: Response of strain energy ($E_s$) under uniaxial strain (S) applied along Y axis in hydrogenated and fluorinated graphene and BN sheet.

Fig-3: Comparison of first derivative of strain energy ($dE_S/dS$) with uniaxial strain (S) in graphene, BN sheet and their hydrogenated/fluorinated counter parts.

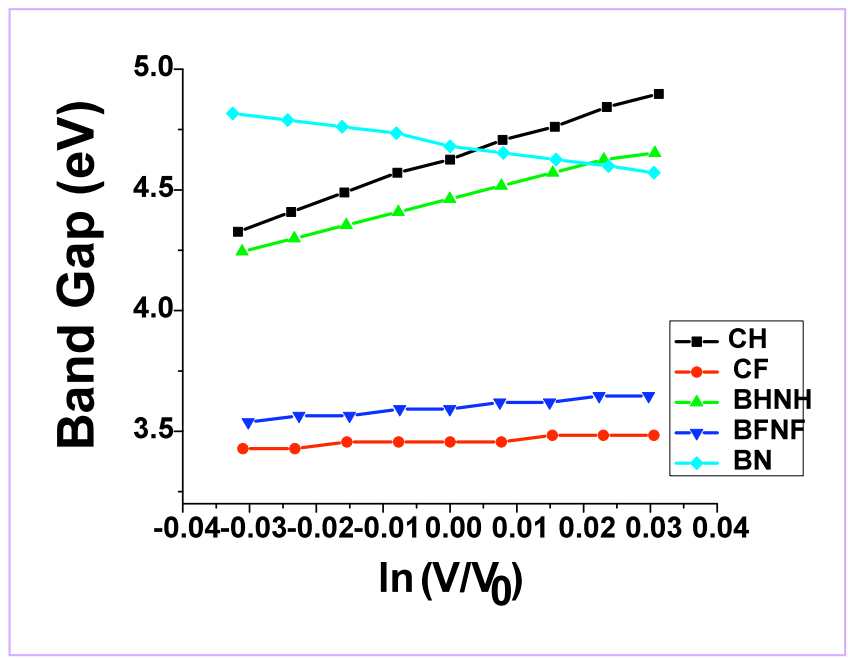

Fig-1

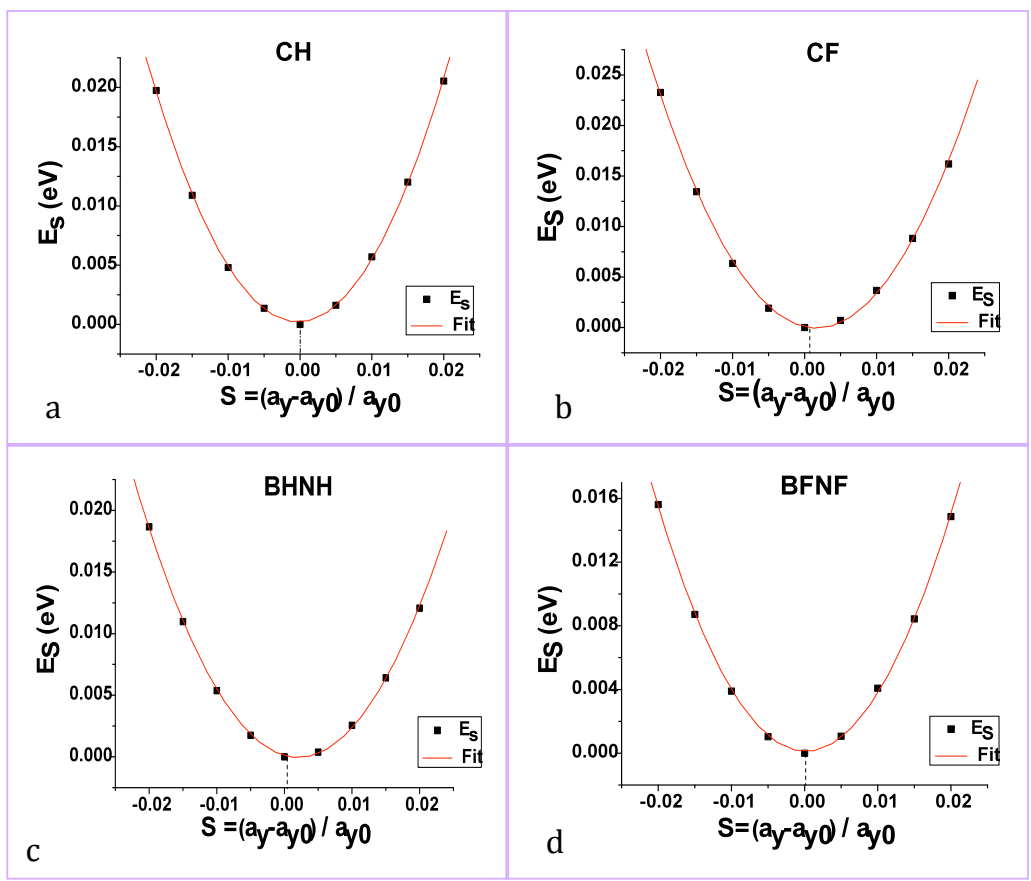

Fig-2

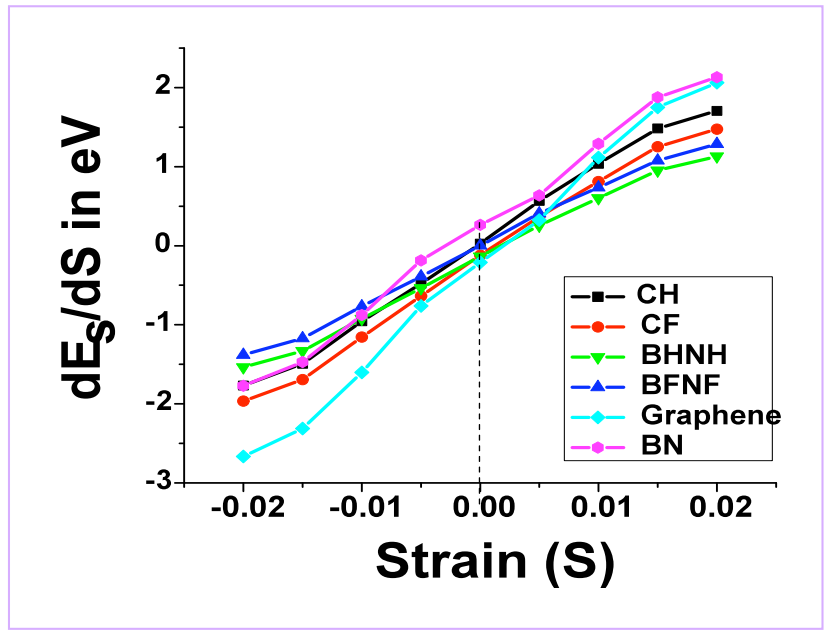

Fig-3

References:
1. A. K. Geim and K. S. Novoselov Nat. Mater. **6**, 183 (2007).
2. K. S. Novoselov, A. K. Geim, S. V. Morozov, D. Jiang, M.I. Katsnelson, I.V. Grigorieva, S.V. Dbonos and A. A. Firsov, Nature (London) **438**, 197 (2005).
3. C. Lee, X. Wei, J. W. Kysar, and J. Hone, Science **321**, 385 (2008).
4. Q. Zhao, M. B. Nardelli and J. Bernholc, PRB **65**, 144105 (2002).
5. J. O. Sofo, A. S.Chaudhari and G. D. Barber, Phys. Rev. B **75**, 153401 (2007).
6. M. H. F. Sluiter, Y. Kawazoe, Phys. Rev. B **68,** 85410 (2003).
7. D. C. Elias, R. R. Nair, T. M. G. Mohiuddin, S. V. Morozov, P. Blake, M. P. Halsall, A. C. Ferrari, D. W. Boukhvalov, M. I. Katsnelson, A. K. Geim and K. S. Novoselov, Science **323**, 610 (2009).
8. A. K. Singh, E. S. Penev and B. I. Yakobson, ACS Nano **4**, 2510 (2010).
9. L. Ci, L. Song, C. Jin, D. Jariwala, D. Wu, Y. Li, A. Srivastava, Z. F. Wang, K. Storr, L. Balicas, F. Liu and P. M. Ajayan, Nat. Mater. **9**, 430 (2010)
10. Y. H. Lu and Y. P. Feng, J. Phys. Chem. C, **113**, 20841 (2009)
11. M. Topsakal, E. Aktürk, and S. Ciraci, Phys. Rev B **79**, 115442 (2009).
12. A. Nag, K. Raidongia, K. Hembram, R. Datta, U. V. Wagmare and C.N.R. Rao, ACS Nano **4**, 1539 (2010)
13. C. Jin, F. Lin, K. Suenaga and S. Iijima, Phys. Rev. Lett. **102**, 195505 (2009).
14. W. Chen, Y. Li, G. Yu, C. Li, S. B. Zhang, Z. Zhou and Z. Chen, J. Am. Chem. Soc. **132,** 1699 (2010).
15. J. Zhou, Q. Wang, Q. Sun and P. Jena, Phys. Rev. B **81**, 085442 (2010), Y. Wang, Phys. Stat. Sol.: Rapid Research Lett. **4**, 34(2009)
16. A. Bhattacharya, S. Bhattacharya, C. Majumder and G.P. Das, Phys. Stat. Sol.: Rapid Research Lett. **4**, 368 (2010)
17. A. Bhattacharya, S. Bhattacharya, C. Majumder and G. P. Das, PRB, 003400 (2011).
18. D. W. Boukhvalov and M. I. Katsnelson, PRB **78**, 085413 (2008)
19. F. Yavari, C. Kritzinger, C. Gaire, L. Song, H. Gulapalli, T. B. Tasciuc, P. M. Ajayan and N. Koratkar, Small **6**, 2535 (2010).

20. M. Topsakal and S. Ciraci, PRB **81,** 024107 (2010); M. Topsakal, S. Cahangirov and S. Ciraci, App. Phys. Lett. **96**, 091912 (2010).
21. M. Cardona and N. E. Christensen, Phys. Rev. B **35**, 6182 (1987).
22. A. Fransceschetti, Phys. Rev. B **76**, 161301 (2007)
23. P. Hohenberg and W. Kohn, Phys. Rev. B **136**, 864 (1964).
24. W. Kohn and L. Sham, Phys. Rev. **140**, A1133 (1965).
25. (a) B. J. Delley, Chem. Phys. **92**, 508 (1990). (b) B. J. Delley, Chem. Phys. **113**, 7756 (2000).
26. J. P. Perdew and Y. Wang, Phys. Rev. B **45**, 13244 (1992).
27. J. P. Perdew, J.A. Chevary, S. H. Vosko, K. A. Jackson, M. R. Pederson, D. J. Singh and C. Fiolhais, Phys. Rev.B **46**, 6671 (1992).
28. H. J. Monkhorst and J. D. Pack Phys. Rev. B **13**, 5188 (1976).